\documentclass[a4paper,USenglish]{article}

\usepackage{arxiv}

\usepackage[utf8]{inputenc} 
\usepackage[T1]{fontenc}    
\usepackage{hyperref}       
\usepackage{nicefrac}       
\usepackage{microtype}      

\usepackage{booktabs}       
\usepackage{amsthm}
\usepackage{amsmath}
\usepackage{mathbbol}
\usepackage{amsfonts}
\usepackage{amssymb}
\usepackage{graphicx}
\usepackage{algorithm}
\usepackage{url}
\usepackage{floatflt}
\usepackage{mathptmx}
\usepackage{epsfig}
\usepackage{wrapfig}
\usepackage{color}

\usepackage{algorithm}
\usepackage[noend]{algpseudocode}
\usepackage{comment}
\usepackage{braket}
\usepackage{pgfplots}
\usepackage{pgf,tikz}
\usetikzlibrary{arrows}

\pgfplotsset{compat=1.14}

\newcommand{\R}{\mathbb{R}}

\newcommand{\eps}{\varepsilon}

\newcommand{\metricspace}{\mathcal{M}}

\newcommand{\diam}{\mathrm{diam}}
\newcommand{\pointset}{P}

\newcommand{\dist}{\delta}

\newcommand{\complexity}{C_{\dist}}

\newcommand{\ignore}[1]{}

\newcommand{\myparagraph}[1]{\textbf{#1.}}

\newtheorem{theorem}{Theorem}
\newtheorem{lemma}[theorem]{Lemma}

\title{Metric Spaces with Expensive Distances}

\author{Michael Kerber\\
Graz University of Technology \\
Institut für Geometrie \\
Kopernikusgasse 24, 8010 Graz, Austria \\
\And
Arnur Nigmetov\\
Graz University of Technology \\
Institut für Geometrie \\
Kopernikusgasse 24, 8010 Graz, Austria \\
}

\begin{document}
\maketitle

\begin{abstract}
In algorithms for finite metric spaces, it is common
to assume that the distance between two points
can be computed in constant time, and complexity
bounds are expressed only in terms of the number
of points of the metric space.
We introduce a different model where we assume
that the computation of a single distance is
an expensive operation and consequently, the goal is
to minimize the number of such distance queries.
This model is motivated by metric spaces
that appear in the context of topological data analysis.

We consider two standard operations on metric spaces,
namely the construction of a $1+\eps$-spanner
and the computation of an approximate nearest
neighbor for a given query point.
In both cases, we partially explore the metric space
through distance queries and infer lower and upper bounds
for yet unexplored distances through triangle inequality.
For spanners, we evaluate several exploration strategies
through extensive experimental evaluation.
For approximate nearest neighbors, we prove that our
strategy returns an approximate nearest neighbor
after a logarithmic number of distance queries.
\end{abstract}


\section{Introduction}

Given a set $\pointset:=\{p_1,\ldots,p_n\}$ of $n$ points
in a metric space $(\metricspace,\dist)$, 
consider the following standard operations:

\begin{description}
\item[Approximate Nearest Neighbor] Given $\eps>0$ and a point $q\in\metricspace$,
find $p_i\in\pointset$ such that, for all $j=1,\ldots,n$,
\[\dist(q,p_i)\leq(1+\eps)\dist(q,p_j)\]

\item[Spanner] Given $\eps>0$, compute a weighted graph $G$ with vertices in $\pointset$
such that for any $u,v\in\pointset$, the shortest path distance between $u$ and $v$
is at most $(1+\eps)\dist(u,v)$.
\end{description}

The performance of algorithms for these problems depends on the number of points,
the dimension of the metric space, and the cost $\complexity$ of computing a distance in the metric space.
It is a common assumption to assume $\complexity$ to be a constant; 
There are good reasons for that: the most common case of a metric space
is $\metricspace=\R^d$ with $d$ some constant, in which case $\complexity$ can be evaluated in $O(d)=O(1)$ time.
Even if $d$ is considered non-constant,
it can always be assumed that $d\leq n$, hence $\complexity$ is at most $O(n)$.
Another typical assumption is that all pairwise distances are part of the input
in which case $\complexity$ is $O(1)$.

However, we argue that in some situations, distance computations
in $\metricspace$ can be costly and $\complexity$ might be incomparable
with $n$. Our motivation comes from topological summaries
such as persistence diagrams~\cite{elz-topological} or Reeb graphs~\cite{reeb-survey}, which are of interest
in the field of topological data analysis. A persistence diagram
is a point set in $\R^2$, and the distance between two diagrams
is determined by a min-cost matching between the point sets.
If the diagrams have $N$ points, computing this matching requires
polynomial time in $N$, and $N$ might well be larger than $n$, the number
of diagrams considered (\cite{cohen2007stability}). For the case of Reeb graphs, the situation is even
worse: while several metrics on Reeb graphs have been proposed (\cite{bauer2014measuring}, \cite{de2016categorified},
\cite{di2016edit}),
not even an constant-factor approximation algorithm is known that runs
in polynomial time in the size of the graphs.
Another instance is a collection of high-resolution images
endowed with the Wasserstein (or Earth Movers) metric (\cite{rubner2000earth}).

In such situations with expensive distance computations, 
it makes sense to study a different cost model, where only the number of distance computations
is taken into account. For instance, that means that quadratic time operations in terms of $n$
are not counted towards the time complexity, as long as these operations do not query any distance
in $\metricspace$. We also ignore the space complexity in our model.

We will restrict to the case of \emph{doubling spaces}, that is, the doubling dimension
of $\metricspace$ is bounded by a constant. 
In that situation, standard constructions from computational geometry provide partial answers:
Using net-trees~\cite{hm-fast}, we can construct a $\eps$-well-separated pair decomposition (WSPD)~\cite{CK-decomposition} using $O(n\log n)$ distance queries; a WSPD in turn yields
an $\eps$-spanner immediately. Net-trees can also be used to compute approximate nearest neighbors
performing $O(\log n)$ distance computations per query point.
Krauthgamer and Lee \cite{krauthgamer2005black} investigated \textit{black box model},
and proved that ANN search for $\eps < 2/5$ can be done efficiently (i.e., in polylogarithmic time, with polynomial preprocessing
and space) if and only if the dimension is $O(\log \log n)$; their bounds count the number of distance computations.
However, for our relaxed cost model, we pose the question whether simpler constructions achieve
comparable, or even fewer distance computations.

We also propose a slight variant of our model: we assume that we also have access to an (efficient)
$2$-approximation algorithm for the distance queries. Queries to this approximation algorithm
are not counted in the model, hence we can assume that for each pair of points $(u,v)$, we
know a number $A_{u,v}$ with $\dist(u,v)\leq A_{u,v}\leq 2\dist(u,v)$. This induces an approximate ordering
of all distances in the metric space, and it is plausible to assume that such an ordering will simplify
algorithmic tasks on metric spaces, at least in practice.

\myparagraph{Contributions}
We propose simple algorithms for spanner construction and approximate nearest neighbor search
and evaluate them theoretically and experimentally in the defined cost model.

\begin{figure}[h]
\centering
\includegraphics[width=6cm]{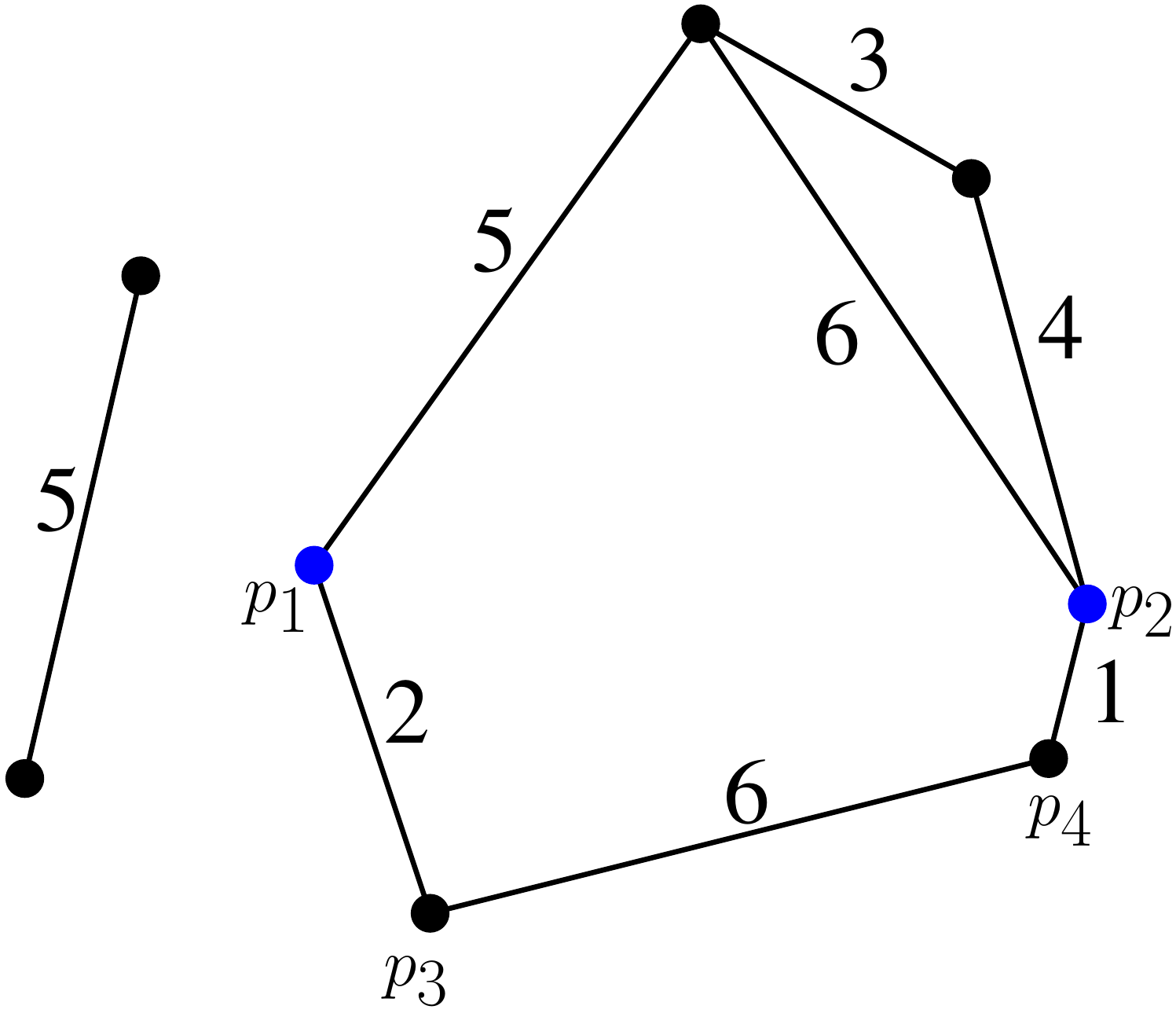}
\caption{The compute distances are shown as edges in a graph. Note that the exact distance
of $p_1$ and $p_2$ is unknown. The shortest path from $p_1$ to $p_2$ has length $9$, which clearly
constitutes an upper bound on the distance by triangle inequality.
However, we can also infer that $\dist(p_1,p_2)\geq 3$:
otherwise, the path from $p_3$ to $p_4$ via $p_1$ and $p_2$
would be shorter than the edge $(p_3,p_4)$, again contradicting
triangle inequality.}
\label{fig:1st_example}
\end{figure}

Our algorithms are based on the following simple idea: since distance computations are expensive
and should be avoided, we try to obtain maximal information out of the distances that have been computed
so far. 
This information consists of lower and upper bounds for unknown distances, obtained from known distances
by triangle inequality (see Figure~\ref{fig:1st_example}). We remark that updating these bounds involves $\Omega(n^2)$ arithmetic
operations whenever a new distance has been computed, turning the method useless in the standard computational model.

We propose several heuristics of how to explore the metric space to obtain accurate lower and upper bounds
with a small number of distance computation. Once the ratio of upper and lower bound is at most $(1+\eps)$
for each point pair, the set of all computed distances forms the spanner.
The experimentally most successful exploration strategy that we found is to
repeatedly query the distance of a pair with the worst ratio of upper and lower bound.
We call the obtained spanner the \emph{blind greedy spanner}, as opposed to the well-known
\emph{greedy spanner} that precomputes all pairwise distances and only maintains upper bounds (\cite{althofer1993sparse}).
Remarkably, we were not able to improve the quality when knowing initial $2$-approximations of all point pairs.
We also compare with a spanner construction based on WSPD. Our simple algorithms tend to give much smaller
spanners on the tested example. Nevertheless, we leave the question open whether our construction
yields a spanner of asymptotically linear size.

For approximate nearest neighbor, we devise a simple randomized incremental algorithm and show that
the number of distance queries to find an approximate nearest neighbor is $O(\log n)$ in expectation.
Our proof is based on the well-known observation that the nearest neighbor changes $O(\log n)$ times
in expectation when traversing the sequence of points, combined with a packing argument certifying that
only a constant number of distances needs to be computed in-between two minima.
We also experimentally evaluate our approach and observe that the approach follows 
roughly the theoretical prediction.

\section{Background and Definitions}
\myparagraph{Doubling dimension}
A metric space is called \textit{doubling} with \textit{doubling constant} $k$,
if every ball of radius $r$ can be covered by at most $k$ balls of radius $r/2$,
and $k$ is the smallest number having that property.
The \textit{doubling dimension} of a doubling space is defined as $\log k$
(since we usually ignore multiplicative constants, the base of the logarithm is not really important; however,
we always use $\log$ to denote the logarithm with base 2).
It is easy to see that a subspace of a space with doubling dimension $d$ 
is always doubling and has the doubling dimension $O(d)$ (but not necessarily $d$).

We shall need the following lemma, which is just a reformulation of the well-known
packing lemma for doubling spaces (see \cite{smid_2009}, Sect. 2.2).

\begin{lemma}
\label{lem:packing_lemma}
 Let $(\metricspace,\dist)$ be a metric space of doubling dimension $d$, and let $P$ be a subset of a ball 
 $B(x,R)$ in $\metricspace$ such that the distance between any two distinct points of $P$ is at least $r$.
 Then 
 \[|P|\leq \left(\frac{4R}{r}\right)^{d}\]

\end{lemma}
\begin{proof}
We can cover $B(x,R)$ with $2^d$ ball of radius $R/2$, each of these balls we can cover with $2^d$
balls of radius $R/4$, etc. Repeating this process $m := \lceil \log \frac{R}{r/2} \rceil$ times, 
we cover
$B(x, R)$ with $2^{md}$ balls of radius at most $r/2$. Since a ball of radius $r/2$ can
contain at most one point from $P$, 
\[|P|\leq 2^{md}= 2^{\lceil \log \frac{R}{r/2} \rceil d}\leq 2^{(1 + \log \frac{R}{r/2})d}=\left(\frac{4R}{r}\right)^{d}.\qedhere\]
\end{proof}

In the following, we will assume throughout that every considered metric space
has a constant doubling dimension.

\myparagraph{Well-separated pair decomposition}
Given $t > 1$, two disjoint subsets $A, B$ of a metric space $(\metricspace, \dist)$ are called $t$-\textit{well-separated},
if 
\[
\forall a \in A \,\, \forall b \in B \,\, \dist(a, b) \geq t \max(\diam(A), \diam(B))
\]
A well-separated pair decomposition (WSPD) is a set of unordered pairs of sets $\{ \{A_1, B_1 \}, 
\dots, \{A_s, B_s\} \}$ such that each pair $\{A_i, B_i\}$ is $s$-well-separated, and for every unordered pair $\{a, b\}$ of distinct points of $\metricspace$ there exists a unique $j$ such that $a \in A_j$ and $b \in B_j$.
The notion of WSPD was introduced by Callahan and Kosaraju \cite{cal-kos-wspd} for Euclidean spaces.
Har-Peled and Mendel \cite{hm-fast} 
introduced the notion of net-trees and
generalized the results of \cite{cal-kos-wspd} for WSPD, proving the following:
\begin{enumerate}
    \item A net-tree for a metric space with $n$ points can be constructed in $2^{O(\mbox{dim})} n \log n$ expected time.
    \item If $\{ \{A_1, B_1 \}, 
\dots, \{A_s, B_s\} \}$ is an $\eps / 16$-WSPD on $\metricspace$, and $a_i \in A_i, b_i \in B_i$ for $i = 1\dots s$
are chosen arbitrarily, then we get an $\eps$-spanner by taking $s$ edges $(a_i, b_i)$.
    \item For $\eps \in (0, 1]$,  an $\eps$-WSPD of size $n \eps^{-O({\dim})}$ can be constructed in $2^{O({\dim})} n \log n + n \eps ^{-O({\dim})}$ expected time. The algorithm uses the net-tree structure.
\end{enumerate}
The algorithm of constructing a net-tree is complicated and not easy to implement. Beygelzimer
et al. \cite{cover-trees} introduced the notion of a cover tree, which is a simpler data structure 
than net-trees. We mention in passing that cover trees can also be used for building
a spanner (this can be proven with the same methods), and we use cover trees for
building WSPD spanners in one of our implementations.

\section{Algorithms for spanner construction}

\myparagraph{Spanners and known constructions}
Let $(\metricspace, \dist)$ be a finite metric space with $n$ points. 
One way to encode the metric space is a complete weighted graph on $\metricspace$,
where the weights correspond to the distances of the points.
A subgraph $G$ of this graph is called a \emph{$(1+\eps)$-spanner} for $(\metricspace,\dist)$ 
if for any pair of points $(u,v)$,
the shortest path distance $d_{uv}$ of $u$ and $v$ in $G$ satisfies $d_{u,v}\leq (1+\eps)\dist(u,v)$.
In other words, the shortest path metric of $G$ is a good approximation of the actual distance for every pair of points.
Clearly, it is a necessary condition that $G$ is connected, hence every spanner must have at least $n-1$ edges.

The \emph{greedy spanner}(\cite{althofer1993sparse}) is a simple algorithm to compute linear-sized spanners:
\begin{algorithmic}
\label{alg:greedy_spanner}
\Function{GreedySpanner}{$P, \eps$}
    \State {$E\gets\emptyset$}
    \State {Sort all pairwise distances of points in $P$}
    \ForAll {pairs $(p_i,p_j)$ in increasing order}
    \State {$d_{ij}\gets$ Shortest path distance in $(P,E)$}
    \If{$d_{ij}>(1+\eps)\dist(p_i,p_j)$}
    \State {Add weighted edge $(p_i, p_j, v)$ to $E$}
    \EndIf
    \EndFor
    \Return {$(P,E)$}
\EndFunction
\end{algorithmic}

The greedy spanner is guaranteed (\cite{althofer1993sparse}) to return a spanner of size $O(n)$
(for constant doubling dimension and fixed $\eps>0$); in experimental study \cite{farshi2009experimental}
it was also shown to return the sparsest graph. However, it clearly
has to compute all $\binom{n}{2}$ pairwise distances in the sorting step;
this means that in our cost model, the greedy spanner has the worst possible
performance.

On the other hand, spanner constructions based on WSPD only compute
$O(n\log n+n \eps^{-d})$ distances 
to construct an $(1+\eps)$-spanner in doubling dimension $d$.
The spanner size is $O(n\eps^{-d})$. Assuming $\eps$ and $d$ again as constants,
this construction yields a $O(n)$-size spanner using only $O(n\log n)$ distance
computations. However, the algorithm is significantly more involved.

\myparagraph{Blind spanners}
We introduce a new framework for constructing spanners
which we call \emph{blind spanners}: the idea is to maintain,
for every pair of points $(p_i,p_j)$,
a lower bound $a_{ij}$ and an upper bound $b_{ij}$ for $\dist(p_i,p_j)$,
initially set to $[0,\infty)$. While there exists some pair for which $\frac{b_{ij}}{a_{ij}}>(1+\eps)$,
we pick one of them, compute its distance and update the lower and upper bounds of
all pairs with respect to the newly acquired information. Here is the pseudocode:

\begin{algorithmic}
\label{alg:blind_spanner}
\Function{BlindSpanner}{$P, \eps$}
    \State {$E \gets \emptyset$}
    \State {$a_{i,j} \gets 0$ for all $1 \leq i,j \leq n$}
    \State {$b_{i,j} \gets \infty$ for all $1 \leq i,j \leq n, i \neq j$}
    \While {$\exists i \neq j : b_{i,j} / a_{i,j} > 1 + \eps$}
    \State {$(i,j) \gets $} \Call{GetNextEdgeToAdd()}{}
    \State {$v \gets \dist(p_i, pj)$}
    \State {Add weighted edge $(p_i, p_j, v)$ to $E$}
    \State \Call{UpdateBounds}{$i, j, v$}
    \EndWhile
\EndFunction
\end{algorithmic}

In this pseudocode we adopt the convention that a positive number divided by 0 is $\infty$
and $\infty$ is larger than any real number,
thus making the predicate in the while loop well-defined.  

We give the details of the \textsc{UpdateBounds} procedure next.
Suppose that
$\dist(p_i,p_j)=v\in\R$ has been computed.
First, we reset $a_{i,j}$ and $b_{j,i}$ to $v$, since the distance
of $p_i$ and $p_j$ is exactly $v$.
To update the upper bound of some entry $b_{k,\ell}$,
we observe that the shortest path from $p_k$ to $p_\ell$ might now
go through the new edge. Hence, we update
\[
    b_{k,\ell}\gets \min_{i,j}\{b_{k,\ell},b_{k,i}+v+b_{j,\ell},b_{k,j}+v+b_{i,\ell}\}
\]
Repeating this for all $k,\ell$ yields the updated upper bounds.
Note that this results in $O(n^2)$ arithmetic operations,
but no distance computation.

For the lower bound, we observe that for any $1\leq k,\ell\leq n$,
\[
    v-b_{k,i}-b_{\ell,j}
\]
is a lower bound for $\dist(p_k,p_\ell)$. Indeed, this follows from
the triangle inequality
\[\dist(p_i,p_j)\leq \dist(p_i,p_k)+\dist(p_k,p_\ell)+\dist(p_\ell,p_j)\]
by rearranging terms and plugging in the upper bounds for $\dist(p_i,p_k)$
and $\dist(p_k,p_\ell)$. An analogue bound holds with $i$ and $j$ swapped. 

Moreover, the inequalities
\begin{align*}
  a_{j,\ell} - v - b_{k,i}&\leq \dist(p_k,p_\ell)\\
  a_{j,k} - v - b_{j,i}&\leq \dist(p_k,p_\ell)
\end{align*}
hold by triangle inequality, and the same is true with $i$ and $j$ swapped.
This yields $6$ lower bounds for $\dist(p_k,p_\ell)$, and $a_{k,\ell}$
is updated to the maximum of these six lower bounds and its current value.

\myparagraph{Heuristics}
The last missing ingredient of our algorithm is the procedure \textsc{GetNextEdgeToAdd},
that is, how to select the next distance to be computed. We propose two natural choices

\begin{description}
\item[\textsc{BlindRandom}] Among all pairs $(i,j)$ where $\frac{b_{i,j}}{a_{i,j}}>(1+\eps)$,
we pick one pair uniformly at random
\item[\textsc{BlindGreedy}] Pick the pair $(i,j)$ which maximizes the ratio $\frac{b_{i,j}}{a_{i,j}}$.
If the maximizing pair is not unique, choose among the maximizing pairs uniformly at random.
\end{description}
The idea behind \textsc{BlindGreedy} is that we query an edge for which we know the least,
in that way hoping to gather most additional information about the metric space.
Also, our conventions imply that in \textsc{BlindGreedy} the edges that have $a_{i,j} = 0$ or $b_{i,j} = \infty$
have the highest priority, so the algorithm first ensures that the graph is connected and there are positive
lower bounds for every edge before it will start adding any other edges. Based on this observation,
we also tested variations of the \textsc{BlindRandom} algorithm, where the algorithm
first enforces connectedness and/or lower bounds (i.e., if there are infinite upper bounds,
then the algorithm can only choose one of the corresponding edges, etc).



The next two heuristics assume the existence of a $2$-approximation algorithm for distance
computation. Denoting by $A_{i,j}$ the number satisfying $\dist(p_i,p_j)\leq A_{i,j}\leq 2 d(p_i,p_j)$,
we sort all pairwise distances according to the values $A_{i,j}$.
This yields a roughly sorted sequence of distance, because when $\dist(p_i,p_j)>2\dist(p_k,p_\ell)$,
then $A_{i,j}>A_{k,\ell}$ is guaranteed.
We propose two further heuristics that attempt to make use of this sorted sequence.
\begin{description}
\item [\textsc{BlindQuasiSortedGreedy}] Traverse the pairs in increasing order with respect to $A_{i,j}$.
\item [\textsc{BlindQuasiSortedShaker}] Alternates between pairs with small and large $A_{i,j}$
by traversing in increasing order of $A_{i,j}$ in odd iterations and in decreasing order
in even iterations.
\end{description}

\textsc{BlindQuasiSortedGreedy} tries to mimic the greedy spanner and hence appears as a natural
choice at first sight. However, anticipating the experimental results, the heuristic yields very poor
results. The reason is that no pair acquires useful lower bounds when only short distance are queried
(the greedy spanner does not have this issue because it knows the distance and hence does not need
lower bounds).
Generally speaking, short distances are good for sharp upper bounds, whereas long distances are
useful for lower bounds. This motivates \textsc{BlindQuasiSortedShaker}
which alternates between short and long distances.

\section{Experiments on spanners}
We run experiments on the points sampled from the low-dimensional Euclidean space to investigate
experimentally the performance of these heuristics. Clearly, for this metric space, our
cost model is not meaningful since distance comparisons are cheap;
but we picked this environment for controlled experiments.
In order to test the \textsc{BlindQuasiSorted} algorithms we multiply the true distance by a factor from $[1,2]$ chosen uniformly.

\begin{figure}[ht]
    \begin{centering}
\begin{tikzpicture}
    \begin{axis}[xlabel=\# Points,ylabel=\# Edges,
        legend pos=outer north east]

    \addplot+ table [x=n_points, y=edges_quasi_sorted_greedy, col sep=comma] {all_methods_dim_2_eps_0.1.txt};
    \addlegendentry{Blind quasi-sorted greedy}

    \addplot+ table [x=n_points, y=edges_quasi_sorted_shaker, col sep=comma] {all_methods_dim_2_eps_0.1.txt};
    \addlegendentry{Blind quasi-sorted shaker}

    \addplot+ table [x=n_points, y=edges_wspd, col sep=comma] {all_methods_dim_2_eps_0.1.txt};
    \addlegendentry{WSPD}

    \addplot+ table [x=n_points, y=edges_blind_random, col sep=comma] {all_methods_dim_2_eps_0.1.txt};
    \addlegendentry{Blind random}

    \addplot+ table [x=n_points, y=edges_blind_random_lower_bound_first, col sep=comma] {all_methods_dim_2_eps_0.1.txt};
    \addlegendentry{Blind random, lower bound first}

    \addplot+ table [x=n_points, y=edges_blind_greedy, col sep=comma] {all_methods_dim_2_eps_0.1.txt};
    \addlegendentry{Blind greedy}

    \addplot+ table [x=n_points, y=edges_greedy, col sep=comma] {all_methods_dim_2_eps_0.1.txt};
    \addlegendentry{Greedy}

\end{axis}
\end{tikzpicture}
\end{centering}
    \caption{Number of edges in blind spanners generated by different variants of the blind
    algorithm. Greedy non-blind algorithm and WSPD algorithm are included for comparison. The plot is for normally distributed points
    in dimension 2, $\eps = 0.1$.}
    \label{fig:spanner_sparseness}
\end{figure}
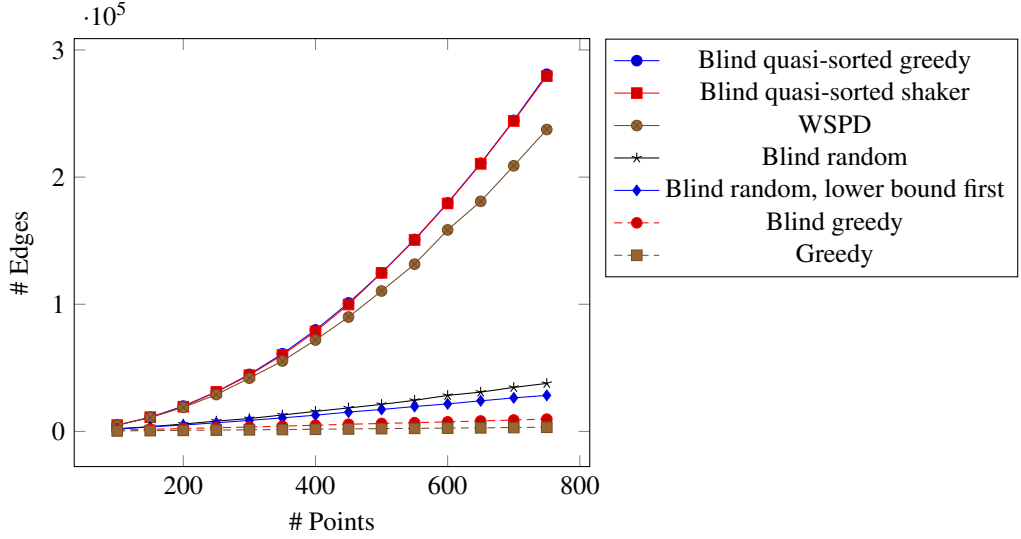

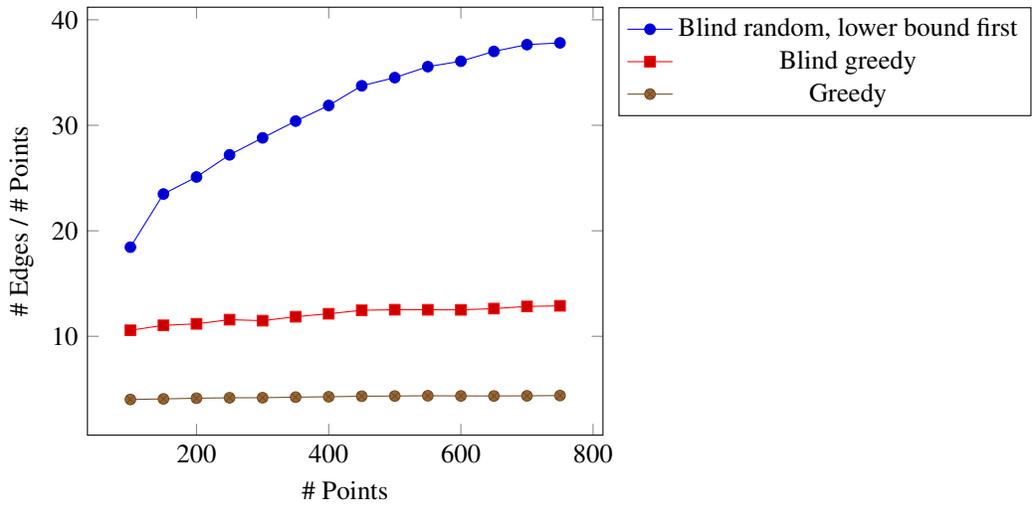
\begin{figure}[ht]
    \begin{centering}
\begin{tikzpicture}
    \begin{axis}[xlabel=\# Points,ylabel=\# Edges / \# Points,
        legend pos=outer north east]





    \addplot+ table [x=n_points, y=ratio_blind_random_lower_bound_first, col sep=comma] {all_methods_dim_2_eps_0.1_ratio.csv};
    \addlegendentry{Blind random, lower bound first}

    \addplot+ table [x=n_points, y=ratio_blind_greedy, col sep=comma] {all_methods_dim_2_eps_0.1_ratio.csv};
    \addlegendentry{Blind greedy}

    \addplot+ table [x=n_points, y=ratio_greedy, col sep=comma] {all_methods_dim_2_eps_0.1_ratio.csv};
    \addlegendentry{Greedy}

\end{axis}
\end{tikzpicture}
\end{centering}
    \caption{Ratio \# edges / \# points for different variants of spanner 
    algorithms. The plot is for normally distributed points
    in dimension 2, $\eps = 0.1$.}
    \label{fig:spanner_ratio}
\end{figure}

\begin{figure}[ht!]
    \includegraphics[width=0.7\textwidth]{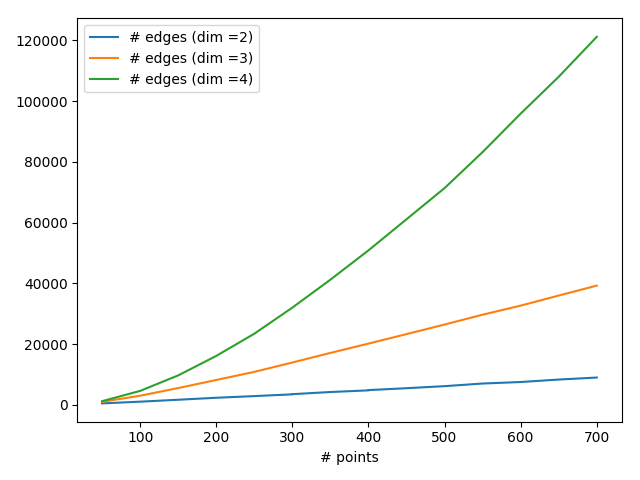}
    \caption{Results of blind greedy spanner for different dimensions.}
    \label{fig:blind_greedy_only}
\end{figure}

\begin{figure}[ht!]
    \begin{centering}
\begin{tikzpicture}
    \begin{axis}[xlabel=\# Points,ylabel=\# Edges,
        legend pos=outer north east]

    \addplot+ table [x=n_points, y=edges_blind_random, col sep=comma] {all_methods_dim_2_eps_0.1.txt};
    \addlegendentry{Blind random}

    \addplot+ table [x=n_points, y=edges_blind_random_connect_first, col sep=comma] {all_methods_dim_2_eps_0.1.txt};
    \addlegendentry{Blind random, connect first}

    \addplot+ table [x=n_points, y=edges_blind_random_lower_bound_first, col sep=comma] {all_methods_dim_2_eps_0.1.txt};
    \addlegendentry{Blind random, lower bound first}

    \addplot+ table [x=n_points, y=edges_blind_random_connect_first_lower_bound_first, col sep=comma] {all_methods_dim_2_eps_0.1.txt};
    \addlegendentry{Blind random, connect first, lower bound first}

\end{axis}
\end{tikzpicture}
\end{centering}
    \caption{Comparison of the four variants of \textsc{BlindRandom} algorithm.}
    \label{fig:blind_rbr_variants}
\end{figure}
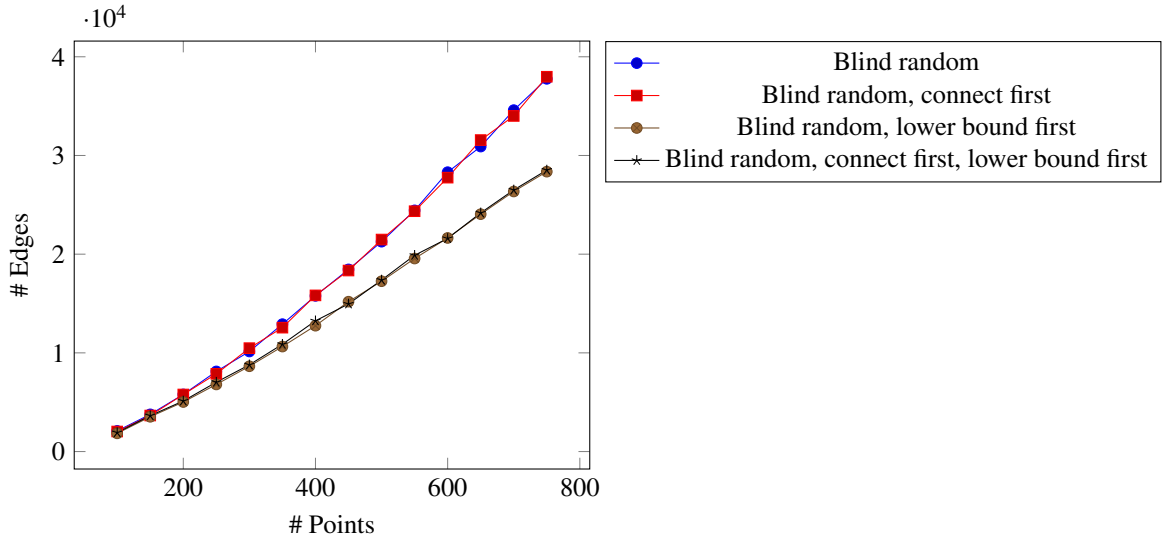


\begin{figure}[ht!]
    \begin{centering}
\begin{tikzpicture}
    \begin{axis}[xlabel=epsilon,ylabel=\# Edges,
        legend pos=outer north east]

    \addplot+ table [x=Epsilon, y=edges_greedy_dim_2, col sep=comma] {eps_dep.csv};
    \addlegendentry{Greedy, dim = 2}

    \addplot+ table [x=Epsilon, y=edges_blind_greedy_dim_2, col sep=comma] {eps_dep.csv};
    \addlegendentry{Blind greedy, dim = 2}



\end{axis}
\end{tikzpicture}
\end{centering}
    \caption{Number of edges in the blind greedy and greedy spanners for different values of $\eps$. Data is for 400 normally distributed points in $\R^2$ and $\R^3$.}
    \label{fig:spanner_eps_dependence}
\end{figure}
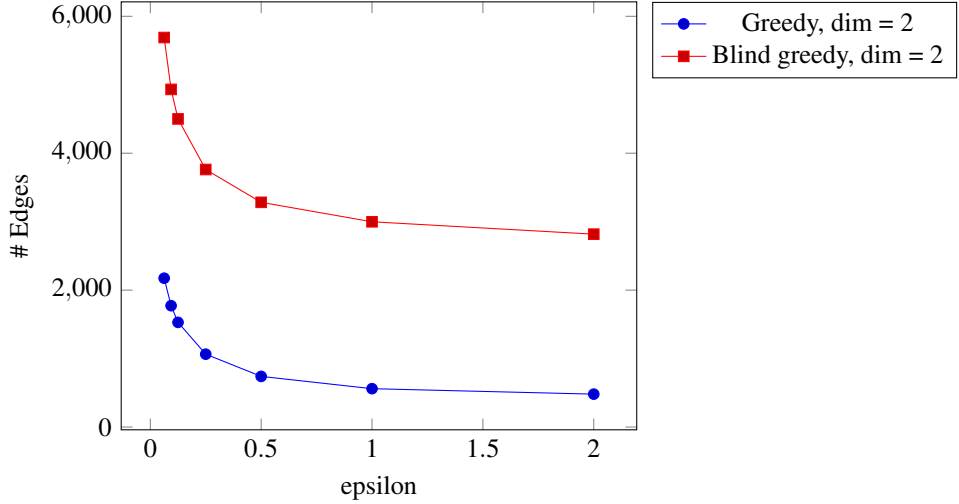


We tested the algorithm for $\eps \in \Set{0.01, 0.1, 0.2, 0.5}$ on the following sets of points in dimensions $d = 2,3,4,5$:
\begin{enumerate}
    \item In the \textbf{uniform} test set points are sampled uniformly at
        random from the unit cube in $\mathbb{R}^d$.
    \item In the \textbf{normal} test set points are sampled from the standard
        normal distribution in $\mathbb{R}^d$.
    \item In the \textbf{clustered} test set we first sample cluster centers uniformly 
        at random from $[0,10000]^d$, and then we add normally distributed noise around
        each of the centers. The number of clusters is chosen so that each cluster
        contains 50 points.
    \item The test set \textbf{exp} consists of points of the form $(2^{\xi_1}, \dots, 2^{xi_d})$,
        where $\xi_i$'s are i.i.d. random variables with uniform distribution on $[1,25]$.
\end{enumerate}
In all experiments the algorithms that we tested compared in the same way,
so we only present results for the \textbf{uniform} point set in dimension 2.

Figure \ref{fig:spanner_sparseness}
shows the number of edges of the spanner for various variants of 
blind and non-blind spanner constructions.
Note that for all blind spanner variants, the number of computed distances
is equal to the spanner size, while for the non-blind greedy spanner,
this number is always $\binom{n}{2}$ and for WPSD it is lower bounded
by the size of the spanner.
We can see that, even though none of the blind spanners can produce
spanners of the same quality (i.e., sparse) as the standard greedy algorithm,
\textsc{BlindGreedy} and all variants of \textsc{BlindRandom}
perform significantly better than both variants of \textsc{BlindQuasiSorted}.
Figure \ref{fig:spanner_ratio} shows the ratio of the number of edges to the number of points.
The ideal behavior is demonstrated by the non-blind greedy spanner,
for which this ratio stays practically constant, confirming the linear growth.
None of the blind algorithms seems to have this property, but among them
the blind greedy spanner is the best one. If we assume that the number of edges
is proportional to $n^\alpha$, then we can try to estimate $\alpha$ by 
linear regression (after taking $\log$).  We give in the table \ref{tbl:regr_coeff_spanner}
the estimated exponents $\alpha$ for \textsc{BlindGreedy} and standard
greedy algorithms. Note that even for the greedy algorithm these estimated
exponents can be significantly larger than 1,
which is explained by the fact that the number of points
on which we computed spanners is not large enough to clearly see
the linear dependence.

\begin{table}[]
\begin{tabular}{|l|l|l|}
\hline
dimension & \textbf{Greedy (non-blind)} & \textbf{Blind greedy} \\ \hline
2         &         1.08                &  1.12                 \\ \hline
3         &         1.24                &  1.41                \\ \hline
4         &         1.42                &  1.77                 \\ \hline
\end{tabular}
\caption{Estimated exponents in the $|E|= C |V|^\alpha$ dependence of the number of edges
on the number of points. The data is for $\eps = 0.1$ and for uniform points.}
\label{tbl:regr_coeff_spanner}
\end{table}

As for different variants of the \textsc{BlindRandom} algorithm,
we note that their performance is almost the same, and the algorithm works
significantly better than \textsc{QuasiSorted} variants, but obviously
worse than the blind greedy variant. There is a consistent, though small,
difference between the variants that do not force lower bounds first and
the other two variants of the \textsc{BlindRandom} (see Figure \ref{fig:blind_rbr_variants}).

WSPD spanners performed poorly in our experiments
on non-clustered data, while the plots in the extensive experimental study \cite{farshi2009experimental}
show that WSPD spanners are very sparse, outperformed only by the greedy algorithm.
We implemented two versions of  WSPD: one for the Euclidean case,
using quadtrees and the algorithm from \cite{hp-book}, and WSPD for general metric
spaces with cover trees (using the base $\tau=1.3$). 
They both give similar results, and we can only conclude
that the advantage of WSPD shows up on larger point sets than the ones we deal with.
The paper \cite{farshi2009experimental} contains experiments for up to 30000 points,
and our blind algorithms, which have at least cubic complexity in the number of points,
are infeasible for such $n$.

We also tested higher dimensions and show the results for the best algorithm, \textsc{BlindGreedy},
in dimensions 2, 3 and 4 in the plot \ref{fig:blind_greedy_only}.
We can see that already in dimension 4, it produces a graph with roughly $\frac{1}{2}\binom{n}{2}$
edges for $700$ points, which clearly shows some degrading for higher dimensions. Still,
we remark that the WSPD spanner remains worse also in the higher-dimensional setup.

The plot in Figure \ref{fig:spanner_eps_dependence}
compares the \textsc{BlindGreedy} and \textsc{Greedy} algorithms
on \textbf{uniform} point sets for different choices of $\eps$. We can see that dependence on $\eps$
is approximately the same for both algorithms. Since it is not cleary seen from the picture, we also note 
that the ratio of the number of edges decreases for
smaller values of $\eps$: 
for $\eps = 2$ the blind greedy spanner contains almost 6 times more edges than the greedy spanner,
while for $\eps = 1/32$ the ratio is 2.6

Summing up, we can conclude from the experiments that 
 the \textsc{BlindGreedy} algorithm performs rather well, but also \textsc{BlindRandom}
algorithm reduces the amount of computed distances substantially, especially if we enforce having non-zero lower bounds
first. 
If the goal is to reduce the number of distance computations, these method seem to be more suitable than a WSPD spanner.
Since the linear spanner size of WSPDs does not show up in the experiments because of the relatively small values of $n$
tested, the experiments are not conclusive regarding the asymptotic size of the blind spanners.
Another noteworthy fact is that quasi-sorted variants produce spanners which are much closer to the complete graph
(\textsc{BlindQuasiSortedGreedy} is worse, requiring all the edges). It would seem plausible that, if we have
access to approximate value of the distance, we could exploit this in the spanner construction, but we could not
find a working heuristic.

\section{Approximate nearest neighbors}
\label{sec:ann}
We consider the standard problem of finding an approximate nearest neighbor: given
$n$ points $P = \Set{p_1, \dots, p_n}$, a query point $q$ and a real number $\eps > 0$,
find $p_i$ such that $\dist(p_i, q) \leq (1 + \eps) \min_{k} \dist(q, p_k)$. This notation will be fixed throughout this 
section, and we shall also use the shorthand notation
\[
    r_i := \dist(p_i, q).
\]
We assume for simplicity
that all exact pairwise distances $\dist(p_i, p_j)$ are already computed
(a slight modification of the algorithm can also be applied if only a spanner is available).
Our goal is to reduce the number of computed distances $\dist(p_i, q)$. 

Our approach can be summarized as follows. 
Fix a random permutation of the points of $P$ 
and consider the points in that order 
(to simplify notation,
we re-index them, so the order is again $p_1, \dots, p_n$).
During the loop, we maintain lower bounds of each $p_i$
to the query point $q$, which are initially all set to $0$.
We also remember the closest neighbor $c$ that we have seen so
far and its distance $v$ to $q$.
We refer to the point $c$ as the \emph{candidate}.
We maintain the invariant that $c$ is an approximate nearest neighbor
to $q$ for the points $\{p_1,\ldots,p_i\}$.
When reaching the point $p_i$, we check whether the lower
bound $a_i$ satisfies $a_i\geq \frac{v}{1+\eps}$.
If so, $c$ remains an approximate nearest neighbor and
we do not query the distance of $p_i$ to $q$.
Otherwise, we compute $\dist(p_i,q)$ and update the lower bounds
of all points according to the newly computed distance.
If $p_i$ is closer to $q$ than $c$, we update $c$ and $v$
accordingly. At the end of the loop, $c$ is an approximate nearest
neighbor. The pseudocode of the procedure follows.

\begin{algorithmic}
\label{alg:ann_blind}

\Function{ApproximateNearestNeighbor}{$P, q, \eps$}
    \State {$[p_1, \dots, p_n] \gets \mbox{random permutation of }P$}
    \State { $a_i \gets 0$ for $i=1,\ldots,n$}
    \Comment {$a_i$ is lower bound for $\dist(p_i, q)$}
    \State {$ c \gets p_1, \quad v \gets \dist(p_1, q)$}
    \Comment {$c$ keeps the current candidate}
    \State \Call {UpdateBounds}{$p_1, v$}
    \For{$i = 2\dots n$}
        \If {$a_i \geq \frac{v}{1+\eps}$}
            \State {\textbf{continue}}
        \Else
            \State {Compute $r_i = \dist(p_i, q)$}
            \State \Call{UpdateBounds}{$p_i, r_i$}
            \If {$r_i < v$}
                \State {$c \gets p_i,\quad v\gets r_i$}
            \EndIf
        \EndIf
    \EndFor
    \State \Return {$c, v$}
\EndFunction
\end{algorithmic}

We remark that we obtain an exact nearest neighbor algorithm
when setting $\eps$ to $0$, which means 
replacing the condition in the if-statement of the loop
with $a_i\geq v$.

The procedure to maintain the lower bounds $a_i$
is very simple and follows directly
from triangle inequality.

\begin{algorithmic}
\Procedure{UpdateBounds}{$p_i, r_i$}
    \For{$k = i + 1, \dots, n$}
        \State {$a_k \gets \max(a_k, |\dist(p_i, p_k) - r_i|)$}
    \EndFor
\EndProcedure
\end{algorithmic}

\begin{theorem}
\label{thm:ann_bound}
    If $(\metricspace, \dist)$ is a doubling space, then, for any fixed $\eps > 0$ the
    algorithm computes $O(\log n)$ distances $\dist(p_i, q)$ in expectation.
\end{theorem}

Towards the proof, we will use the following geometric lemma which
can be summarized as follows: if $\dist(p_i,q)$ is computed in the algorithm,
further distance computations of points very close to $p_i$ or very far from $p_i$
will be avoided.

\begin{lemma}
\label{lem:bound_lemma}
Assume $r_i=\dist(p_i,q)$ is computed in the algorithm, and let $j>i$.
\begin{enumerate}
\item If $\dist(p_i,p_j)\geq (1+\frac{1}{1+\eps}) r_i$, the algorithm will not compute the distance
of $p_j$ to $q$.
\item If $\dist(p_i,p_j)\leq\frac{\eps}{1+\eps}r_i$, the algorithm will not compute
the distance of $p_j$ to $q$.
\end{enumerate}
\end{lemma}
\begin{proof}

\definecolor{uququq}{rgb}{0.25,0.25,0.25}
\definecolor{xdxdff}{rgb}{0.49,0.49,1}
\definecolor{qqqqff}{rgb}{0,0,1}
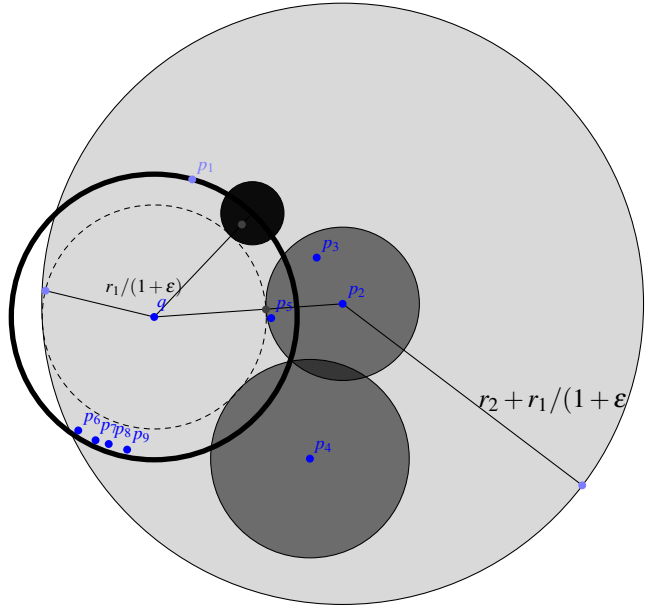
\begin{figure}[ht!]
    \centering
\begin{tikzpicture}[line cap=round,line join=round,>=triangle 45,x=0.5cm,y=0.5cm]
\clip(-1.5,-7.5) rectangle (17,9.2);
\draw [fill=black,fill opacity=0.5] (8.52,0.98) circle (1.02cm);
\draw [fill=black,fill opacity=0.15] (8.52,0.98) circle (4cm);
\draw [fill=black,fill opacity=0.95] (6.12,3.39) circle (0.42cm);
\draw [fill=black,fill opacity=0.5] (7.65,-3.14) circle (1.32cm);
\draw [line width=2pt] (3.51,0.63) circle (1.9cm);
\draw [dash pattern=on 2pt off 2pt] (3.51,0.63) circle (1.49cm);
\draw (3.51,0.63)-- (0.62,1.33);
\draw (3.51,0.63)-- (6.12,3.39);
\draw (8.52,0.98)-- (3.51,0.63);
\draw (8.52,0.98)-- (14.89,-3.85);
\draw (11.88,-1) node[anchor=north west] {$r_2 +r_1 / ( 1 + \eps$};
\begin{scriptsize}
\fill [color=qqqqff] (3.51,0.63) circle (1.5pt);
\draw[color=qqqqff] (3.71,0.94) node {$q$};
\fill [color=xdxdff] (0.62,1.33) circle (1.5pt);
\draw[color=black] (3.24,1.41) node {$r_1/(1+\eps)$};
\fill [color=qqqqff] (8.52,0.98) circle (1.5pt);
\draw[color=qqqqff] (8.9,1.29) node {$p_2$};
\fill [color=uququq] (6.48,0.83) circle (1.5pt);
\fill [color=qqqqff] (7.83,2.21) circle (1.5pt);
\draw[color=qqqqff] (8.21,2.52) node {$p_3$};
\fill [color=qqqqff] (7.65,-3.14) circle (1.5pt);
\draw[color=qqqqff] (8.01,-2.82) node {$p_4$};
\fill [color=qqqqff] (6.62,0.6) circle (1.5pt);
\draw[color=qqqqff] (6.98,0.92) node {$p_5$};
\fill [color=qqqqff] (1.49,-2.39) circle (1.5pt);
\draw[color=qqqqff] (1.87,-2.08) node {$p_6$};
\fill [color=qqqqff] (1.95,-2.65) circle (1.5pt);
\draw[color=qqqqff] (2.31,-2.33) node {$p_7$};
\fill [color=qqqqff] (2.3,-2.75) circle (1.5pt);
\draw[color=qqqqff] (2.68,-2.43) node {$p_8$};
\fill [color=qqqqff] (2.79,-2.9) circle (1.5pt);
\draw[color=qqqqff] (3.17,-2.57) node {$p_9$};
\fill [color=xdxdff] (14.89,-3.85) circle (1.5pt);
\fill [color=uququq] (5.84,3.09) circle (1.5pt);
\fill [color=xdxdff] (4.52,4.29) circle (1.5pt);
\draw[color=xdxdff] (4.89,4.61) node {$p_1$};
\end{scriptsize}
\end{tikzpicture}
\caption{First two steps of the ANN algorithm. First $p_1$ is 
chosen as the current candidate, and we must compute $\dist(p_2, q)$. After that the algorithm will not compute
distance to any of the points inside the heavily shaded ball or outside the lightly shaded ball that are centered at $p_2$, because their lower bounds allow us to discard them. Note that the point $p_5$, which
is closer to $q$ than $p_1$, also will not be a candidate, and at least one of the points $p_6,p_7,p_8,p_9$ in the annulus between the dashed and solid circle, which are further from $q$ than $p_5$, will be chosen as $c$. This shows that in our algorithm the distance from the candidate to $q$ can drop \textit{slower} than in the bruteforce algorithm, thus Theorem \ref{thm:ann_bound} does not immediately follow from standard backwards analysis. The small black ball between the dashed circle and the solid circle has radius $v_1 \eps / (1 + \eps)$; it is the ball that we use in the packing argument, because it is smaller than any of the lightly shaded balls that correspond to points like $p_2$ and $p_4$, that is, the points that do not improve $v$.}
\label{fig:ann_illustration}
\end{figure}

\begin{figure}[hb!]
    \includegraphics[width=0.7\textwidth]{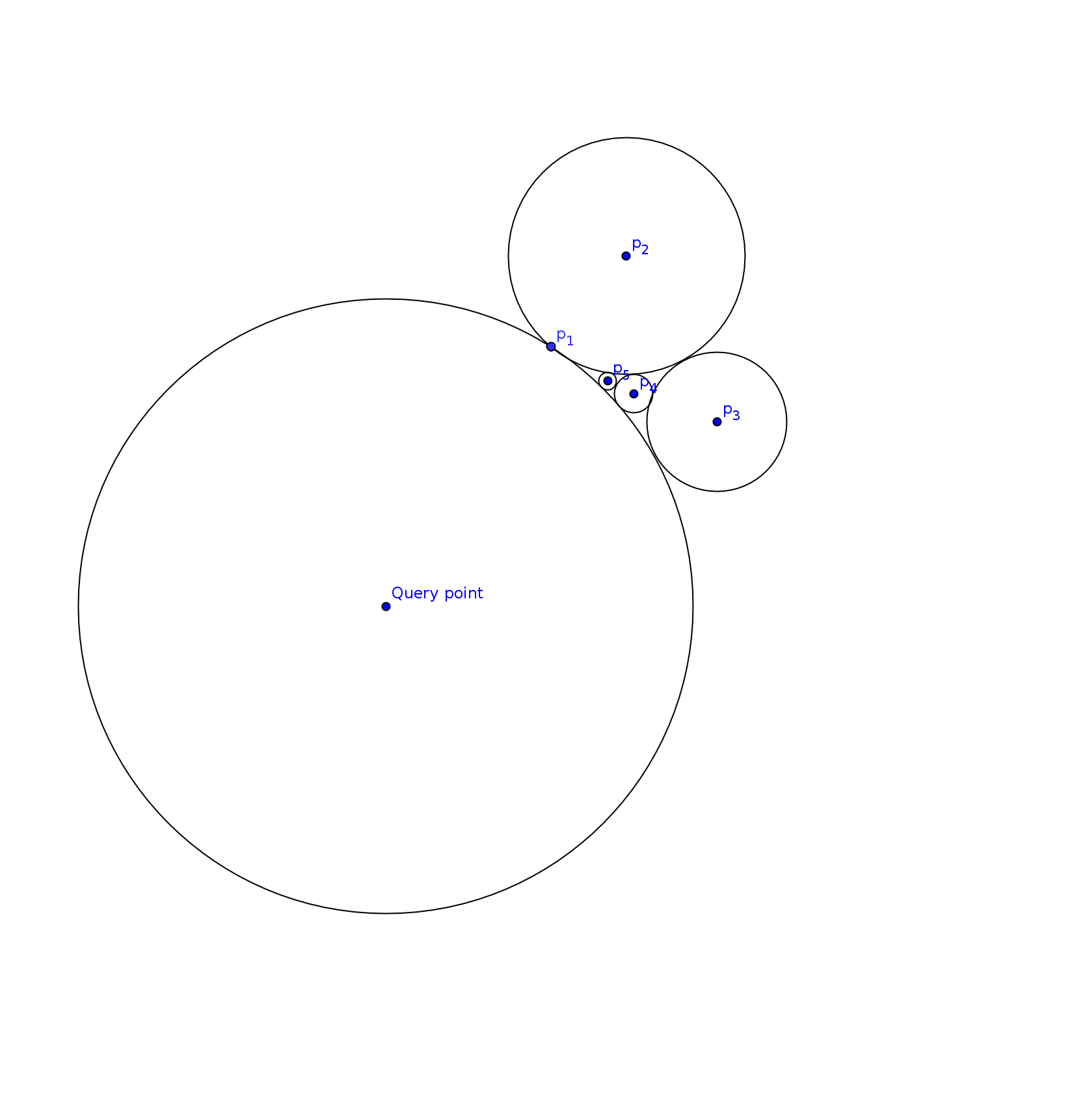}
    \caption{Example of point set where exact nearest neighbor search cannot be accelerated by maintaining bounds.
    The exact nearest neighbor is the point $p_1$, next point $p_i$ is placed 
    in the curvilinear triange formed by the balls around the query point, $p_2$ and $p_{i-1}$. Even verifying that $p_1$
    is the true nearest neighbor cannot be done without computing all distances $\dist(p_i,q)$. Indeed, every computed
    $\dist(p_i,q)$ allows to exclude the region in the corresponding ball around $p_i$, but all these balls contain only one $p_i$.}
    \label{fig:exact_bad_example}
\end{figure}

The algorithm computes $r_i$ by assumption and updates
all lower bounds. For $p_j$, it sets $a_j\gets \max (a_j,|\dist(p_i,p_j)-r_i|)$.
If $\dist(p_i,p_j)\geq (1+\frac{1}{1+\eps})r_i$, it follows that 
\[a_j\geq (1+\frac{1}{1+\eps})r_i - r_i = \frac{r_i}{1+\eps}.\]
Likewise, if $\dist(p_i,p_j)\leq\frac{\eps}{1+\eps}r_i$,
\[a_j\geq r_i-\dist(p_i,p_j) \geq r_i - \frac{\eps}{1+\eps}r_i = \frac{r_i}{1+\eps}.\]
In both cases, after the point $p_i$ is handled,
$v\leq r_i$ clearly holds. Since $v$ is only decreasing
and $a_j$ is only increasing in the algorithm,
it follows that $a_j\geq \frac{v}{1+\eps}$ when $p_j$ is handled,
so the algorithm proceeds without a distance computation.
\end{proof}

In what follows, we let $c_i$ denote the candidate
at the end of the $i$-th iteration of the loop, and $v_i$
the distance to $\dist(c_i, q)$, $i = 1, \dots, n$. Clearly, $v_1,\ldots,v_n$ is a decreasing sequence.
With the previous lemma, we can derive an upper bound for the number
of distance computations in an arbitrary subsequence of $p_1,\ldots,p_n$
as follows. 

\begin{lemma}
\label{lem:sequence_lemma}
Among the points $p_k,\ldots,p_\ell$ with $1\leq k< \ell\leq n$, the algorithm computes at most
\[\left(\frac{4(2+\eps) v_k}{\eps v_\ell}\right)^{d}\]
distances to $q$.
\end{lemma}
\begin{proof}
By the first part of Lemma~\ref{lem:bound_lemma}, every point in $p_k,\ldots,p_\ell$
whose distance to $q$ is queried lies in the ball of radius $(1+\frac{1}{1+\eps})v_k=\frac{2+\eps}{1+\eps} v_k$
around $c_k$. Moreover, if the distance of two points $p_i$ and $p_j$ with $k\leq i<j\leq\ell$
is computed, the second part of Lemma~\ref{lem:bound_lemma} implies that $\dist(p_i,p_j)> \frac{\eps}{1+\eps}r_i\geq \frac{\eps}{1+\eps}v_\ell$.
Hence, all points in $p_k,\ldots,p_\ell$ for which the algorithm computes the distance
have a pairwise distance of at least $\frac{\eps}{1+\eps}v_\ell$. The statement follows by applying Lemma~\ref{lem:packing_lemma}.
\end{proof}

A consequence of the lemma is that as long as a candidate $c$ is fixed in the algorithm,
the number of computed distances is a constant (since $v_k=v_\ell$). 
This means that to prove Theorem~\ref{thm:ann_bound}, it would suffice to show
that the candidate changes only a logarithmic number of times in expectation.
While we have not found a simple proof for this claim, we can prove the statement with a slight variant of
that argument.

\begin{proof} (of Theorem~\ref{thm:ann_bound})
In the sequence $p_1,\ldots,p_n$, let $p_k$ be a point 
such that $r_i<r_k$ for all $1\leq i\leq k-1$. We call an element
of this form a \emph{minimum} of the sequence. 
A standard backwards analysis argument~\cite{seidel-backwards} shows that the probability
of $p_k$ being a minimum is at most $1/k$, so that the number of minima
in the sequence is $O(\log n)$ in expectation.

Note that for $\eps>0$,
a minimum $p_k$ is not necessarily the candidate $c_k$ because a previous point
in the sequence close to $p_k$ might have caused the lower bound $a_k$ to be
in the interval $[\frac{v_k}{1+\eps},v_k]$, which leads to not
computing the distance $r_k$. However, it is true that
$v_k\leq (1+\eps)r_k$, because otherwise, $c_k$ would not be an approximate
nearest neighbor of $\{p_1,\ldots,p_k\}$.

Now, let $p_k$, $p_\ell$ be two consecutive minima in the sequence
(we also allow that $\ell=n+1$ if $k$ is the last minimum in the sequence).
Note that $v_{\ell-1}\geq r_k$ because each $v_j$ is equal to $r_i$ for
some $i\leq j$, and in the sequence $r_1,\ldots,r_{\ell-1}$, $r_k$
is minimal by construction. Using Lemma~\ref{lem:sequence_lemma},
the number of distance computations among the points
$p_k,\ldots,p_{\ell-1}$ is at most
\[
\left(\frac{4(2+\eps) v_k}{\eps v_{\ell-1}}\right)^{d}\leq \left(\frac{4(2+\eps)(1+\eps)r_k}{\eps r_k}\right)^{d}=\left(\frac{4(2+\eps)(1+\eps)}{\eps}\right)^{d},\]
which is a constant depending only of $\eps$ and $d$, irrespective of the length of the sequence.
Since $p_1,\ldots,p_n$ decomposes into $O(\log n)$ such sequences in expectation, the result follows.
\end{proof}

We point out that the proof fails for $\eps=0$ because in that case, we cannot exclude an $\eps$-ball
of close-by points as in the second part of Lemma~\ref{lem:bound_lemma}, and the packing argument fails.
Indeed, as the example in Figure~\ref{fig:exact_bad_example} shows, there are point sets where the expected number
of distance computations for exact nearest neighbor is linear.

Finally, we remark that a fast $2$-approximation algorithm for $\dist$ would lead to a straight-forward
optimization: compute a $2$-approximation of $\dist(p_i,q)$ for all $1\leq i\leq n$
and let $m$ denote the minimal approximate distance encountered. Then, we can discard all points
whose approximate distance is larger than $2m$, and run the above algorithm on the remaining points.

\section{Experiments on approximate nearest neighbors}

In order to experimentally evaluate the performance of our algorithm,
we generate random point sets and random query points, and for each query point
run the algorithm 10 times. The average number of distances to the query point
that were actually computed is the measure that we are interested in.
We average the results over 10 different instances of the point
set and query point in order to see the trend clearer; thus
each point on the plots in this section is the result of averaging of 100 runs
of the code (10 instances, 10 random permutations per instance).

\begin{figure}[ht]
    \includegraphics[width=0.7\textwidth]{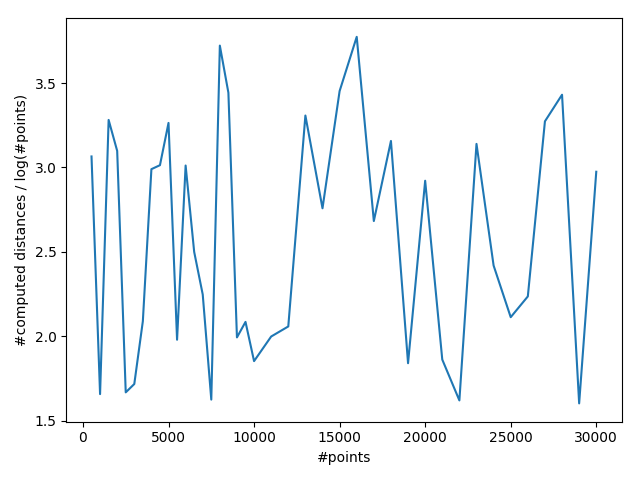}
    \caption{Ratio $\log(\mbox{computed distances}) / n$ for ANN algorithm. Data is for uniformly distributed
    points.}
    \label{fig:ann_const_ratio}
\end{figure}

\begin{figure}[ht]
    \includegraphics[width=0.7\textwidth]{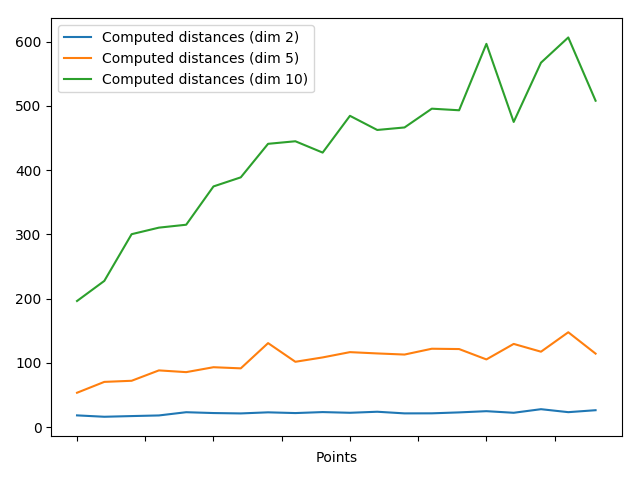}
    \caption{Number of computed distances for different dimensions. Points are chosen uniformly, $\eps = 0.01$.}
    \label{fig:ann_dimension_dependence}
\end{figure}

We used the following methods of generating random points:
\begin{enumerate}
    \item Uniform. Points are sampled uniformly at random from the unit cube in $\R^d$.
    \item Normal. Points are sampled from the normal distribution.
\end{enumerate}
Query points were sampled from the uniform distribution on the cube $[-10, 10]^d$
and from the normal distribution centered at the origin with scale 100,
thus we get query points that are "inside" the point set and also "outside".
We sample data in dimensions up to 20 and for $\eps \in \Set{0.001, 0.005, 0.01, 0.05, 0.1}$,
the maximal number of points is $30,000$.

In order to empirically verify the upper bound $O(\log n)$,
we plot the number of computed distances divided
by the logarithm of the number of points in figure \ref{fig:ann_const_ratio} (for $d = 2$).
We see that this ratio, though fluctuating a lot, remains in the interval $[1,4]$. This not just confirms
the theoretical upper bound, but also shows that the algorithm in the low-dimensional case
really computes only a very small number of distances to the query point.
As expected, in high dimensions the algorithm does not perform as well.
In Figure \ref{fig:ann_dimension_dependence} we plot the average number of computed distances
for $d = 2, 5, 10$. While for $d = 2$ the growth is hardly noticeable, for $d = 10$
the sublinearity of the growth becomes clear only when the number of points is relatively large,
approaching 30000.

\section{Conclusion and future work}
\label{sec:conclusion}
We have introduced a new cost model for the analysis of algorithms
for metric spaces that fits the situation that computing an individual distance
is more costly than other types of primitive operations.
Our theoretical and experimental results are under the usual assumption
that the metric space has a low doubling dimension.
However, in our motivating example of collections of persistence diagrams
or Reeb graphs, this assumption does not hold. For instance,
the space of persistence diagrams has an infinite doubling dimension.
Nevertheless, realistic data sets are usually not just a random sample
in that infinite-dimensional space, but have structures
(e.g. clusters of close-by diagrams) which should be favorable for our approach
We plan to consider the quality of our algorithms for persistence diagrams
as future work.

On the theoretical side, the obvious next question is whether our strategy
for blind spanners yields a linear spanner in expectation. 
Our experiments are not conclusive enough in this respect to make this conjecture yet.
However, it has been brought to our attention\footnote{Yusu Wang, personal communication}
that the size of the blind spanner is bounded by the \emph{weight} of the WSPD
which is the sum of the  cardinalities of all pairs in a WSPD.
The weight of a WSPD can be quadratic, but preliminary
experimental evaluation on worst-case examples do not show such a quadratic
behavior. Therefore, we postpone the theoretical analysis of the spanner construction
to an extended version of this article.

The existence of a $2$-approximation algorithm did not help us to significantly
reduce the number of exact distance computations, although it seems obvious
that knowing the all approximate distances is useful.
We pose the question what heuristic could make more use of this feature.

\bibliography{bib}
\bibliographystyle{plainnat}

\end{document}